# Intense high contrast femtosecond K-shell x-ray source from laser-driven Ar clusters

L. M. Chen[1*], F. Liu[1], W. M. Wang[1], M. Kando[2], X. X. Lin[1], J. L. Ma[1], Y. T. Li[1], S. V. Bulanov[2], T. Tajima[2], Y. Kato[2], Z. M. Sheng[1,3], J. Zhang[1,3#]

[1]*Beijing National Laboratory of Condensed Matter Physics, Institute of Physics, CAS, Beijing 100080, China*
[2]*Advanced Photon Research Center, Japan Atomic Energy Agency, 8-1 Umemidai Kizugawa, Kyoto 619-0215, Japan*
[3]*Department of Physics, Shanghai Jiao Tong University, Shanghai 200240, China*

Bright Ar K-shell x-ray with very little background has been generated using an Ar clustering gas jet target irradiated with an 800 mJ, 30 fs ultra-high contrast laser, with the measured flux of 1.1 x $10^4$ photons/mrad$^2$/pulse. This intense x-ray source critically depends on the laser contrast and the laser energy and the optimization of this source with interaction is addressed. Electron driven by laser electric field directly via nonlinear resonant is proved in simulation, resulting in effective electron heating and the enhancement of x-ray emission. The x-ray pulse duration is demonstrated to be only 10 fs, as well as a source size of 20 μm, posing great potential application for single-shot ultrafast x-ray imaging.

[*] On leave from Japan Atomic Energy Agency. Electronic address: lmchen@aphy.iphy.ac.cn
[#] Electronic address: jzhang@aphy.iphy.ac.cn

Hard x-ray emission from femtosecond (fs) laser produced plasmas has been extensively studied in the past years, for example see refs. [1-5]. Such hard x-ray sources have a number of interesting applications in imaging [6-9]. This kind of intense and ultrafast hard x-ray source can pose as a possible alternative to the synchrotron radiation due to its compactness, its low cost compared to big facility and its fs pulse duration, and is, therefore, of interest for practically using in majority of labs and hospitals.

Unfortunately, there are serious obstacles to limit laser-driven hard x-ray sources which are available nowadays in imaging applications, e.g: the limited x-ray conversion efficiency and the spectrum contrast ratio from Kα to the background [10]. The hard x-ray emission produced by high intensity laser-solid interactions relies on hot electrons [11], which penetrate into the solid target and produce hard x-ray emission via K-shell ionization and bremsstrahlung. Typically, the x-ray continuum in spectrum is strong and usually contains 90% of total x-ray energy with photon energies > 1 keV [4]. In the worst case, the energetic x-ray tail will greatly reduce the subject contrast of in-line radiography [8]. The micro-droplet is another candidate x-ray source material, with even higher conversion efficiency, but the strong x-ray energetic tail limits its application for imaging for the same reason. X-ray sources from laser-cluster interaction, usually using the low intensity and/or the normal contrast laser pulse, suffer from the low conversion efficiency, which is usually $\sim 10^{-7}$ for Ar and Kr clusters [12, 13]. Again, its spectrum also has a strong bremsstrahlung continuum, which reduces the K-shell line emission contrast. Therefore, the laser-driven hard x-ray source with more intense flux and higher spectrum contrast ratio is crucial for imaging application.

In this Letter, we have presented generation of intense Ar K-shell x-rays with very weak continuum background using a small cluster target irradiated with an intense high contrast fs laser pulse. The intensity of the Ar K-shell emission in the forward direction has been measured to be $1.1 \times 10^4$ photons/mrad$^2$/pulse, which is strong enough for single-shot x-ray imaging. Nonlinear resonant is proved to be the predominant mechanism in case of high contrast fs laser interacting with nm-size clusters which driven electron efficiently in 10 fs temporal scenario and enhance x-ray emission.

The experiment are carried out in the Advanced Photon Research Center with the J-Karen laser facility, which is a 10 Hz, 800 mJ Ti:Sapphire laser working at the centre wavelength $\lambda$=*800 nm*. The pulse with duration $\tau_0$=*28 fs* is focused with an f /13 off-axis parabola onto a focal spot of size $w_0$=*16 μm*. In the focal region the laser intensity is $I=1.6x10^{18}$ $W/cm^2$ in average. With the help of OPCPA, the laser pulse contrast compared to its ns prepulse has been improved to $10^9$. A supersonic pulsed gas (Ar) jet is used. This jet is produced by a specially designed pulsed gas valve with a shockwave-free nozzle, which is 3 mm in diameter in the exit. A strong magnet is placed after the nozzle to expel electrons. A filtered 16-bit LCX single photon counting CCD (Roper Scientific) is located in the laser propagation direction to detect the x-ray flux with a photon energy > 0.8 keV. The calibration shows that the resolving power of the CCD as a dispersionless spectrometer in ~ 3 keV is about 20. A knife edge is introduced to measure the x-ray source size. A probe beam is used for detecting the shadowgraph by a visible CCD. The cluster size is estimated by employing the Hagena scaling law [14]. An average size of ~8 nm in diameter is assumed at stagnation pressure 4 MPa.

**Figure 1** shows the Ar x-ray spectrum obtained in a single shot. Compared to the x-ray spectrum from the solid Cu target in similar laser parameters, the Ar K-shell emission shows much higher contrast (signal to continuous background ratio) in the spectrum. The amplitude of the continuum part of the spectrum is about 1% compared to the peak of the K emission line. The integrated K-shell photon number is higher than 95% of all the photons in the spectrum, whereas it is only ~10% in the case of a solid target [4] and previous large-size cluster target [13]. It should be emphasized that the x-ray flux depends critically on the laser contrast. The flux will be reduced 2 orders if the laser pulse contrast decreases from $10^9$ to $10^7$ in the constant laser pulse energy. Pre-expansion of solid-density cluster by the laser pre-pulse leads to un-effective heating for electrons, as described hereinafter, and results in the decrease of x-ray flux in this case. We also need to mention that this high contrast, quasi-monochromatic structure for the K-shell emission would disappear if energetic electrons are generated, unlike the case in ref. [15]. The x-ray continuous background becomes a Bi-Maxwellian distribution [16], dramatically reducing the contrast of the x-ray line emission.

By changing laser and interaction conditions, the Ar x-ray emission is optimized. The x-ray emission increases linearly by increasing the gas backing pressure (P) and then saturates when P > 4MPa, whereas the cluster size is continue to increase. A threshold is observed for x-ray emission as a function of laser energy, as shown in **the inset of Fig. 1**. When the laser energy is higher than 40 mJ, the x-ray flux will increase steeply and reaches a top at 80 mJ, which correspond to laser intensity 1.6x$10^{17}$ W/cm$^2$. The more interesting phenomenon is that the detected photon intensity is a function of the nozzle displacement away from laser focal spot, as shown in **Fig. 2**. If the laser focuses on the edge of plasma, energetic electron will be generated and the K-shell x-ray flux is very unstable. As the nozzle moves away from the laser focal spot, the photon flux decreases at first, gets a minimum and then starts to increase. It reaches to the maximum 1.1x$10^4$ photons/mrad$^2$ when the nozzle moves close to the OAP ~ 6 mm. In this case, the K-shell x-ray generation is very stable. The laser intensity on the edge of plasma is reduced to $10^{16}$ *W/cm*$^2$, which is similar to our previous observation using much smaller laser facility [16]. If the nozzle displacement is less than 5 mm, the electron will be accelerated and produce strong γ-ray background. The plasma volume obtained in the case of best focus and defocus proved that a long and straight laser channel leads to the enhancement of the K-shell emission. As seen the shadowgraph imaging in **Fig. 2**, the plasma channel with diameter 60 μm in the defocus case is much longer than the case of best focus, providing much larger interaction volume. The x-ray source sizes detected via the knife edge technique in the case of 6 mm defocus is ~ 20 μm in FWHM, smaller than the focal spot size.

Simulations using a 2D fully electromagnetic particle in cell (PIC) code have been performed, where a linearly polarized laser pulse with $sin^2$ pulse envelope is launched along the +x direction onto a cluster with 10 nm in diameter and 10 $n_{cr}$ in density. **Figure 3(a)** shows snapshots of the electron distribution profile at various times. At the early stage of the laser pulse, electrons are driven by the laser electric field and quiver along laser polarization direction in each laser period, see also in **Fig. 3(b)**. **Figure 4(a)** shows the electron number of inner electrons decrease depend on time period but for each half optical cycle, part of electrons will return into cluster boundary that reveals the electron quivering procedure. This phenomenon is defined as non-linear resonant (NLR) electron heating mechanism [17]. These electrons, with quiver energy shown as these small spikes on the curve in **Fig. 4(d)**, will pass through the cluster and stimulate k-shell ionization to produce x-rays when the electron energy larger than Ar ionization threshold. We need to mention that the quiver electrons posses non-Maxwellian electron energy in **Fig. 3(b)** which is optimal to stimulate Ar K-shell x-ray with much less continuous background. However, this quiver phenomenon becomes blurry at the pulse latter stage because an electron cloud form by escaping electrons around the cluster and compensate the effect of the laser electric field, as shown in **Fig. 3(a), (c)** and **Fig. 4**. Therefore, we conclude that the K-shell x-ray photon generation arises from electrons quivering in the early stage of intense laser fields. Electrons possess the quiver energy $E_q=e^2E^2/(2m_e\omega_0^2) = 9.3x10^{-14}\ I\lambda^2$ *(eV)*. For $I=10^{17}$ $W/cm^2$, it is ~ 6 keV. This is high enough and in fact optimal to stimulate Ar K-shell photons at energies of ~3 keV, considering the maximized cross section of Ar [18]. It explains that there is a laser threshold intensity to stimulate K-shell x-ray emission in **Fig. 1**. In **figure 4(d)**, only 4-5 optical cycles, with spikes higher than 5 keV, are fit for this energy level which correspond to the duration about 10 fs. Considering the Ar K-shell vacancy lifetime ~ 4.8 fs [19], we conclude that the K-shell x-ray pulse duration as short as 10 fs and demonstrate this is an ultra-short hard x-ray source. On the other hand, the cluster inner electron obtains energy much lower than $E_k$ which leads to no contribution to K-shell ionization in our case, as shown in **Fig. 4(c)**. This rules out the linear resonant (LR) heating [20] and the long-term ionization procedure of cluster in case of using a normal contrast laser.

The transfer from NLR to LR is proved experimentally by using the high contrast laser. By changing the distance between the compressor gratings away from "zero" position at constant laser energy, the incomplete compensation of the accumulated phase nonlinearities results in negatively (positively) chirped pulses having a gentle (steep) rise time [10]. In case of laser irradiation with intensity $1x10^{17}$ $W/cm^2$, the x-ray flux as a function of laser pulse duration in negatively skewed and positively skewed case show different tendency. As shown in **Fig. 5(a)**, the flux of using positively skewed pulse drops much gentle than the case of negatively skewed. Simulations in **Fig. 4(d)** using different pulse shape successfully reproduced this phenomenon. The laser intensity still strong enough to drive electron efficiently when laser pulse duration slightly expands, stimulating the NLR. Positively skewed pulse drives higher spikes than the case of negatively pulse irradiation, even the latter have more weaker spikes which is not strong enough to efficiently stimulate Ar K-shell ionization. However, when the laser intensity drops to $1 \times 10^{16}$ W/cm$^2$, the laser electric field cannot heat electron effectively in the fs time scale. The x-ray flux is dramatically reduced, as shown in **Fig. 5(b)**. In this case, the cluster need some time, usually 500fs-1ps [20], to make it expand to the critical density for stimulating the LR mechanism. Just as expected, the x-ray flux in **Fig. 5(b)** reaches a peak at that laser pulse duration which means, we successfully controlled the cluster heating mechanism from NLR to LR using the high contrast laser.

In summary, we have presented generation of intense Ar K-shell x-rays with very weak background using a cluster target irradiated with an intense fs laser pulse. The intensity of the Ar x-ray emission has been measured to be $1.1 \times 10^4$ photons/mrad$^2$/pulse, which is strong enough for a phase contrast x-ray imaging in single laser shot using a imaging plate 20 cm away from x-ray source [16]. Nonlinear resonant is believed to be the dominant mechanism in case of high contrast fs laser interacting with nm-size clusters which driven electron efficiently in 10 fs temporal scenario and enhance x-ray emission. Together with the measured source size of 20 μm and the emission duration 10 fs, the peak brightness of the radiation is estimated to be ~$2 \times 10^{21}$ photons/s/mm$^2$/mrad$^2$, which is comparable to the peak brightness of the third generation synchrotron radiation sources. This ultra-intense, monochromatic and *fs* hard x-ray source opens the window for "single-shot" laser-driven x-ray ultrafast applications.

This work was supported by the NSFC (Grant No. 60878014, 10675164, 60621063, 10735050 and 10734130), National Basic Research Program of China (973 Program) (Grant No. 2007CB815102) and the National High-Tech ICF program.

# Figure Captions

Figure 1: X-ray spectra of Ar when the laser beam was focused on the cluster target. The x-ray spectrum of Cu with a solid target is also shown (dotted line). The inset shows the photon flux depends on laser electric field.

Figure 2: Channeled emission characteristics: The K$\alpha$ photon emission depends as a function of the target displacement. The inset show the shadowgraph obtained in the case of the best focus (up) and a position +6 mm away from the focal spot (down). The positive displacement means the nozzle moves closer to the OAP.

Figure 3: Snapshots of the electron density distribution (a) of cluster at various times. The energy spectra and phase spectra ($P_x$, $P_y$) of outer electrons at various time are shown on the early stage of laser pulse (b) and on the peak of pulse (c).

Figure 4: Laser pulse shape dependent: The temporal distribution of number of inner electrons (a), energy of total electrons (b), energy of inner electrons (c) and energy of outer electrons (d). The laser pulse with rising/down edge 5/5 (red line), 5/10 (black line), 10/5 (blue line) are used in simulations.

Figure 5: Ar K-shell x-ray emission as a function of laser pulse width with negatively skewed (red line) and positively skewed (black line) for a full energy pulse (a) and low energy pulse (b) irradiation.

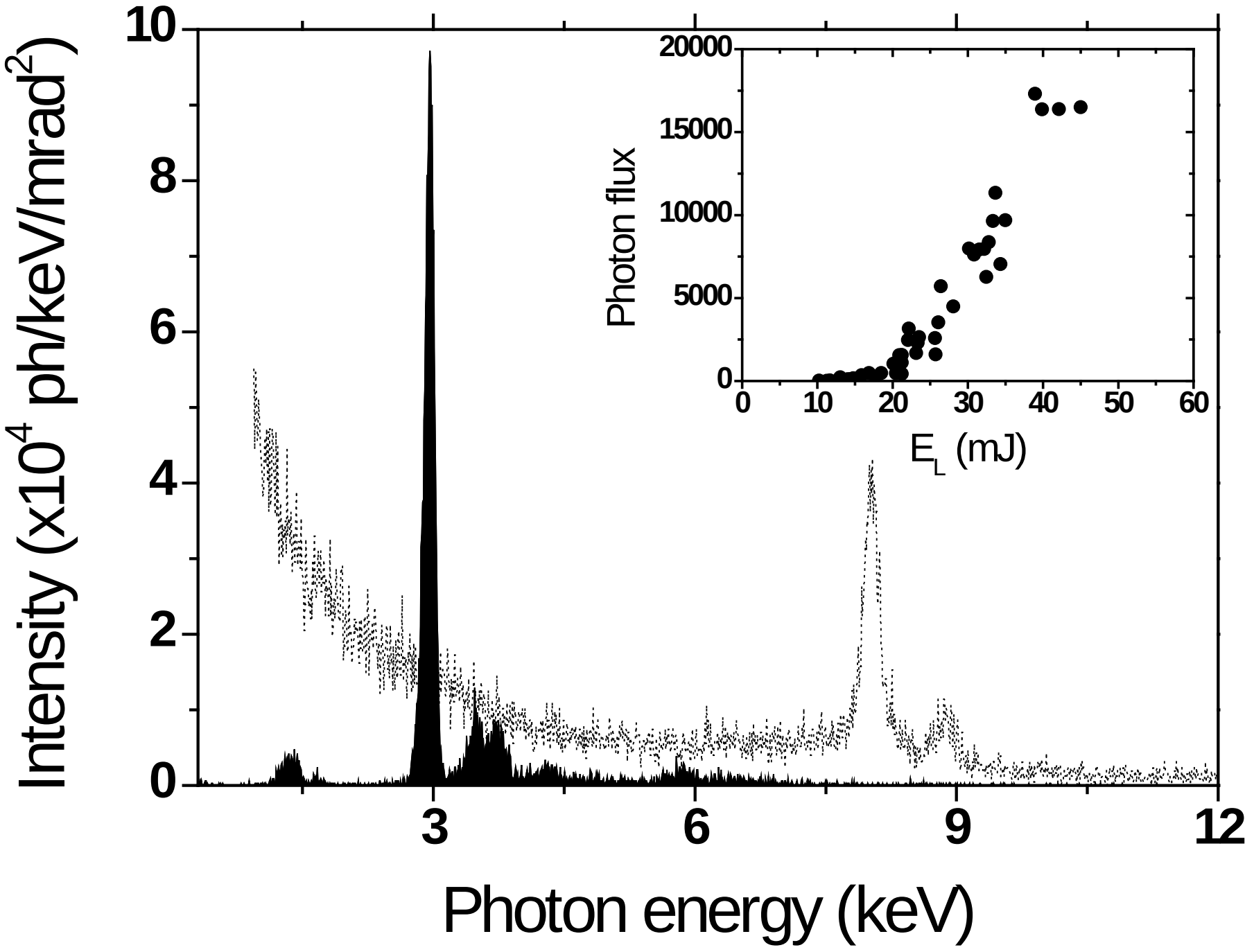

Fig. 1: L. M. Chen et al.

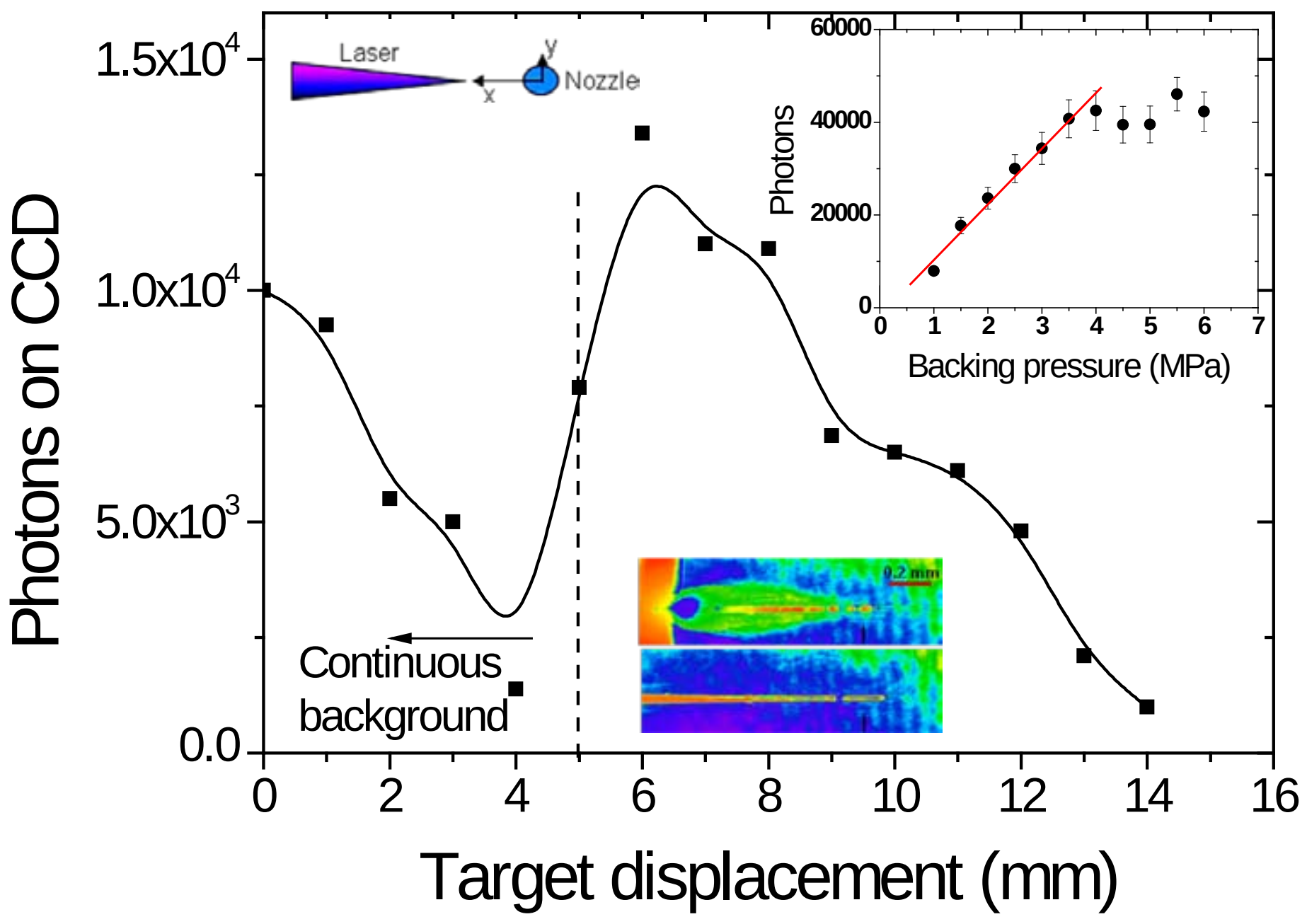

Fig. 2: L. M. Chen et al.

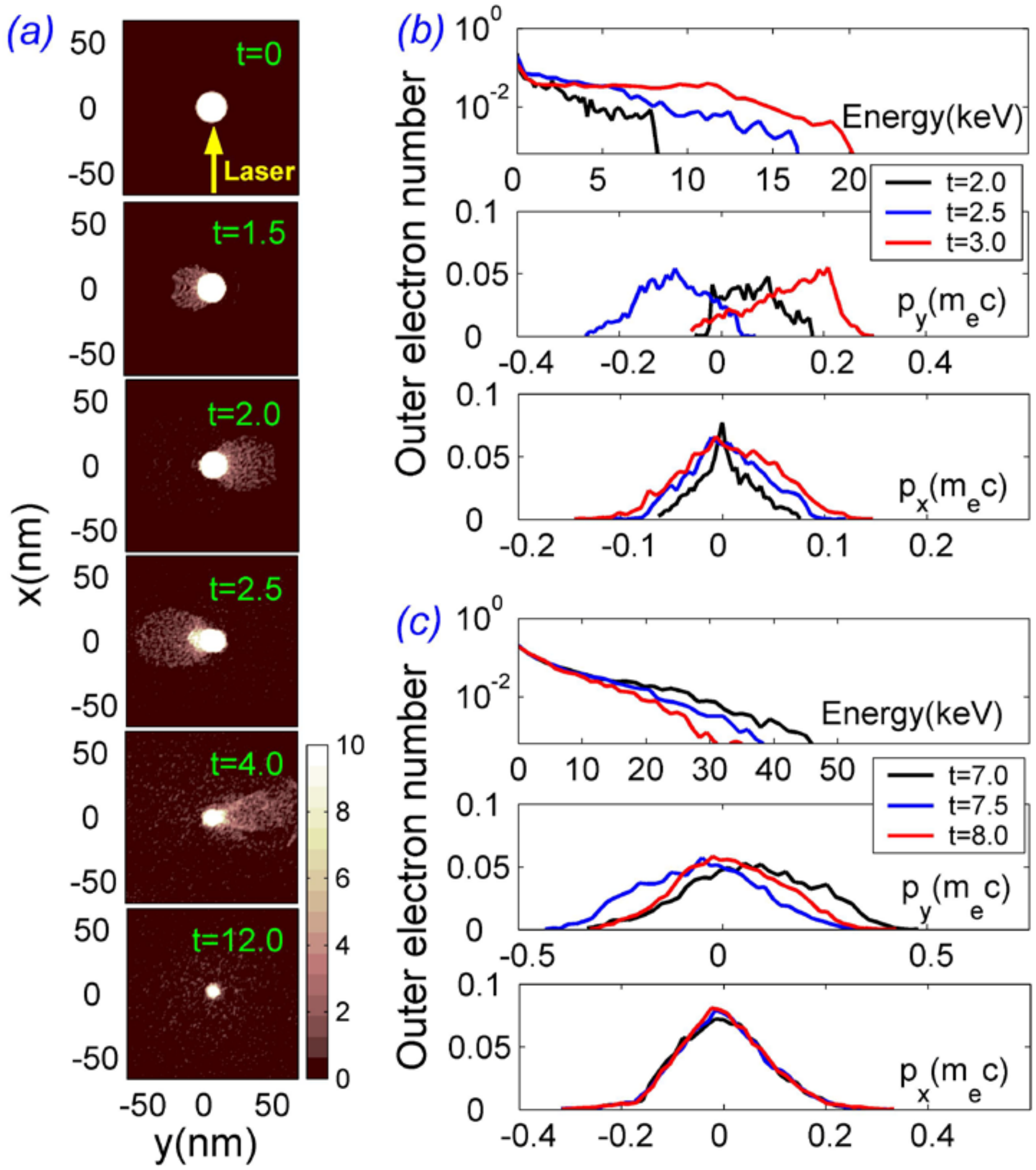

Fig. 3: L. M. Chen et al.

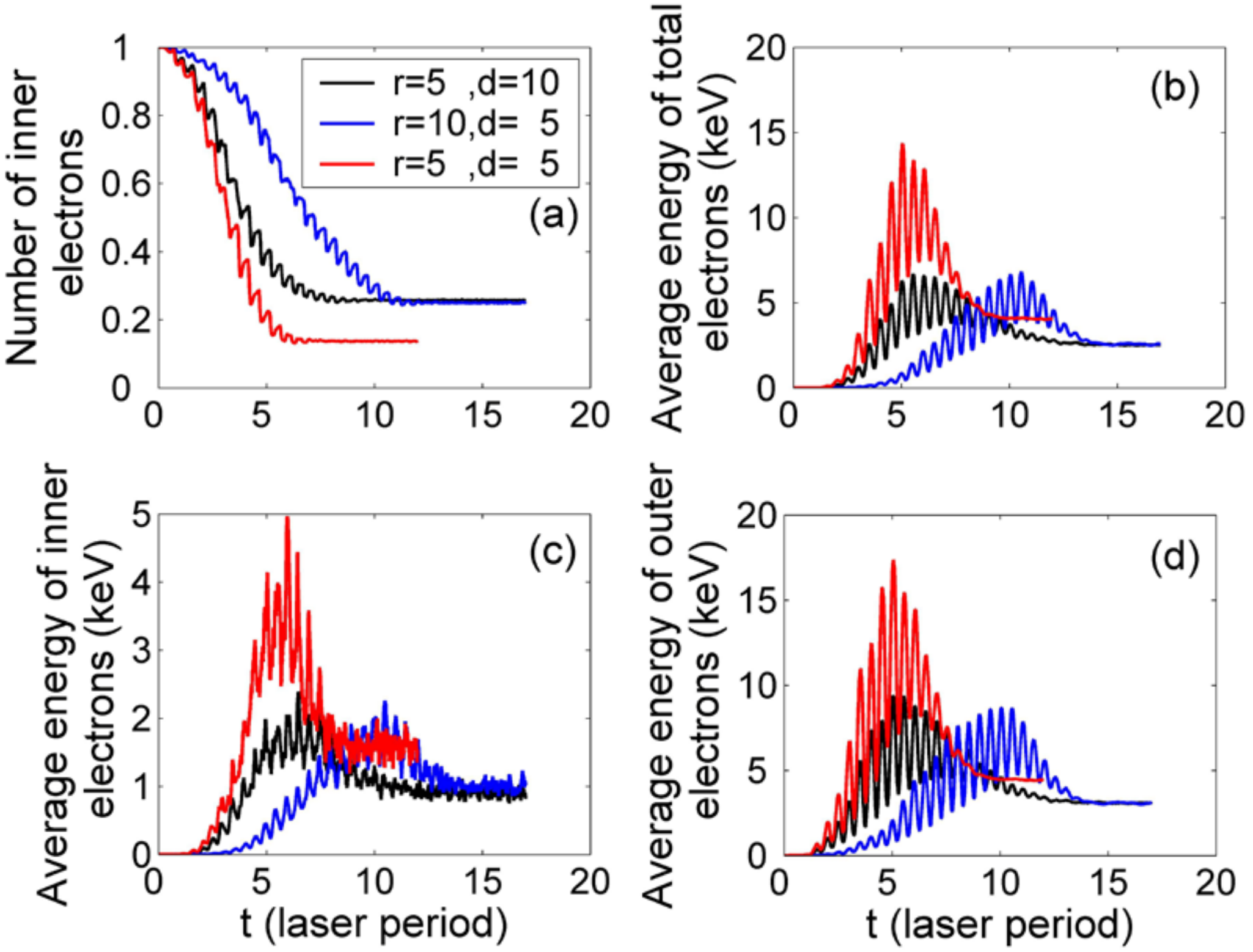

Fig. 4: L. M. Chen et al.

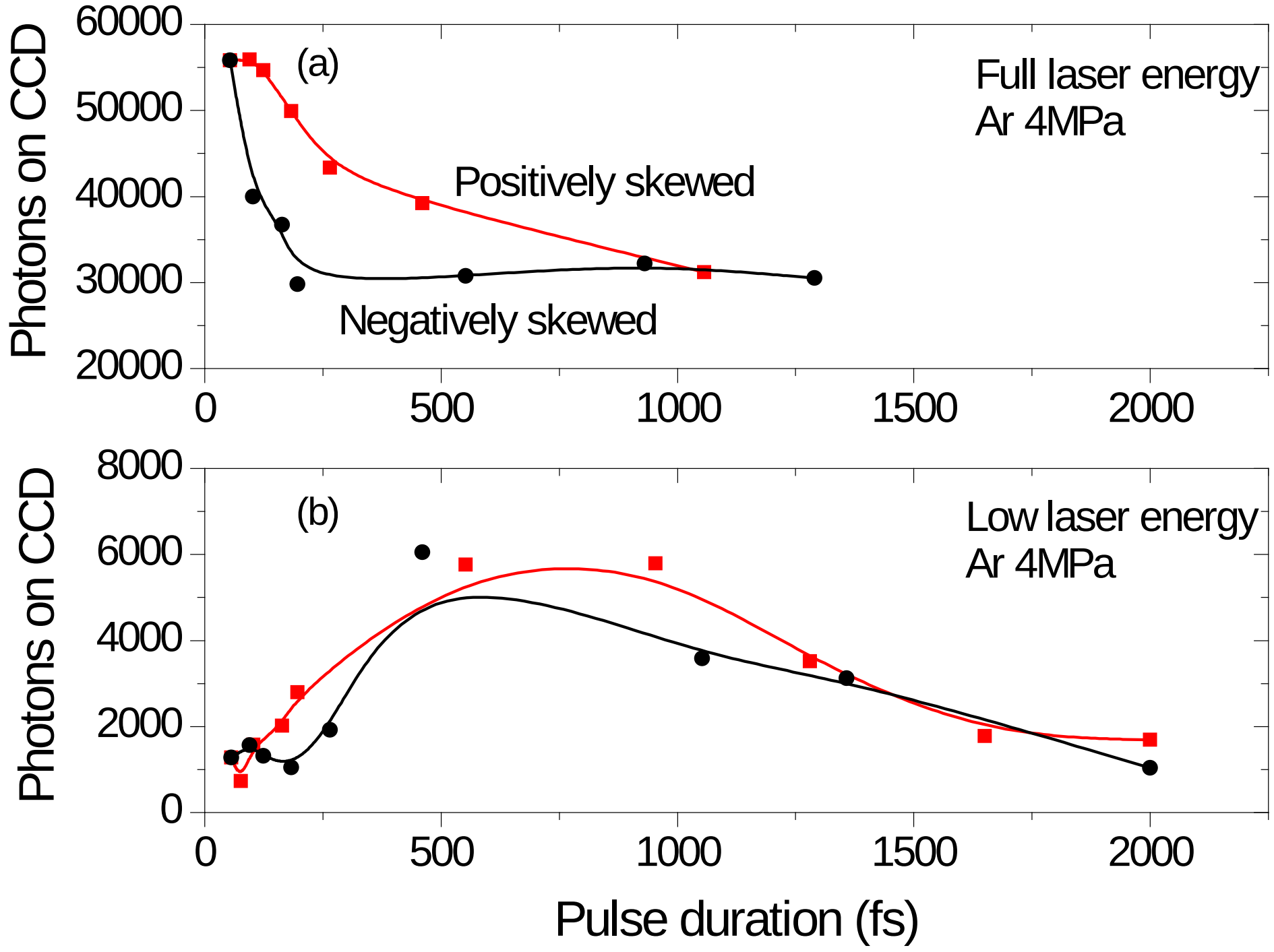

Fig. 5: L. M. Chen et al.